\documentclass[fleqn,usenatbib]{mnras}
\usepackage{newtxtext,newtxmath}
\usepackage[T1]{fontenc}
\DeclareRobustCommand{\VAN}[3]{#2}
\let\VANthebibliography\thebibliography
\def\thebibliography{\DeclareRobustCommand{\VAN}[3]{##3}\VANthebibliography}
\usepackage{graphicx}	
\usepackage{amsmath}	
\usepackage{gensymb}
\title[Pulse modulation due to pulsar-asteroid collision]{Pulse Modulation as a Signature 
of the Asteroid-Neutron Star Collision Model for High-Energy Transients}

\author[Bagchi, Layek, Saini, Sarkar, Srivastava and Godaba Venkata]{
Partha Bagchi $^{1,2}$ \thanks{E-mail: parphy@niser.ac.in, parphy85@gmail.com}, 
Biswanath Layek $^3$ \thanks{E-mail: layek@pilani.bits-pilani.ac.in}, 
Dheeraj Saini $^4$ \thanks{E-mail: 23pph001@lnmiit.ac.in},
Anjishnu Sarkar $^4$ \thanks{E-mail: anjishnu@lnmiit.ac.in}, 
\newauthor
Ajit M. Srivastava $^5$ \thanks{E-mail: ajit@iopb.res.in} and 
Deepthi Godaba Venkata $^3$ \thanks{E-mail: 
p20210075@pilani.bits-pilani.ac.in}\\
\\
$^1$ School of Physical Sciences, National Institute of Science Education 
and Research, Bhubaneswar, India\\
$^2$  (At present) Physics Department, Marwari College; A Constituent Unit of Purnea
University, Kishanganj- 855107, Bihar, India\\
$^3$Department of Physics, Birla Institute of Technology and Science, Pilani, Pilani Campus, 
Vidya Vihar, Pilani, Rajasthan 333031, India \\
$^4$ Physics Department, The LNM Institute of Information Technology, Jaipur-302031, India \\
$^5$ Institute of Physics, Sachivalaya Marg, Bhubaneswar-751005, India
}

\date{Accepted XXX. Received YYY; in original form ZZZ}
\pubyear{\the\year{}}

\begin{document}
\label{firstpage}
\pagerange{\pageref{firstpage}--\pageref{lastpage}}
\maketitle
\begin{abstract}
Asteroid-neutron star collision models have been proposed as possible sources of high-energy transients, 
such as gamma-ray bursts (GRBs) and fast radio bursts (FRBs). The sequence of events following the impact 
of the asteroid and finally dissolving into the neutron star can have several other observable consequences. 
We propose that due to the development of the off-diagonal moment of inertia (MI) components, the 
merger's aftermath can lead to the wobbling of the pulsar (assuming the neutron star happens to be a pulsar). 
Using sample values of various parameters, viz., size, shape, the locations of the deposits, and the pre-existing 
pulsar deformation parameter ($\eta$), we calculate the detailed pulse profile modulation of the pulsar. 
We observe a distinct pattern of pulse profile modulation on a characteristic timescale enhanced by a factor 
of $1/\eta$ compared to the pulse timing.
Importantly, even small changes in the MI components, of order $\epsilon$, can produce large pulse profile 
modulations of order $\epsilon/\eta$ (depending on the relative location of asteroid material deposition).
Thus, if an asteroid-neutron star collision is responsible for a high-energy transient, 
the associated pulse profile modulation may serve as a falsifiable observational signature of such an event.
\end{abstract}


\section{Introduction}
\label{section:sec1}
Before the puzzle could settle following the discovery of sixteen short gamma-ray bursts observed in 
1969-1972 \citep{grb1973}, the detection of a unique gamma-ray burst on 5 March 
1979 \citep{event79-1,event79-2,event79-3} further deepened the mystery in gamma-ray astronomy. Despite decades 
of study, the origin of GRBs remains unresolved. The existing models broadly associate long-duration GRBs with 
the collapse of massive stars \citep{Woosley1993, Kumar2015} and short-duration bursts with compact object 
mergers \citep{Berger2014}. While these scenarios explain a few observed features, various alternative mechanisms 
have also been proposed. These include GRBs powered by the highly magnetized neutron stars \citep{Metzger2011}, 
phase transitions in quark or hybrid stars \citep{Ouyed2002, Bombaci_2004}, and the neutron star-asteroid 
collisions, etc. The collisional impact of asteroids on neutron stars was initiated by 
\citet{Newman_1980,Colg81}, where the authors showed that asteroid-neutron star collisions can reproduce 
several temporal and energetic features of GRBs. A similar mechanism has also been considered for the fast 
radio bursts, millisecond-duration radio transients ({\it Lorimer burst}), presumably of extragalactic 
origin \citep{Lori07,Thor13}. The impact of an asteroid onto a neutron star surface was proposed to trigger 
coherent radio emission through magnetospheric disturbances, offering a natural explanation for the short 
timescales and high brightness temperatures of FRBs \citep{Geng2015, Dai_2016}. Thus, the asteroid–neutron 
star collision model provides a unifying framework by linking two apparently different astrophysical events 
within a common physical scenario.

Beyond the prompt burst emission, such collisions may also have other testable signatures. 
In particular, the deposition of asteroid material can perturb the neutron star's moment of inertia (MI), 
generating MI's off-diagonal components in spheroidal pulsars and, consequently, inducing wobbling motions, 
in addition to any wobbling already present. Pulsar wobbling has been investigated in several contexts, such as 
free precession, internal superfluid dynamics, and magnetospheric asymmetries \citep{Stairs_2000, Jones_2012, Haskell2015} 
and is known to cause 
observable pulse profile modulations \citep{Bagchi_2022}. Recently, \citet{Bagchi_2022} demonstrated that 
such pulse profile modulations could serve as a diagnostic tool for probing transition to exotic QCD phases
hypothesized to exist in the cores of neutron stars. Such a proposal is based on the fact that phase transition-induced 
density fluctuations inside a neutron star's core would lead to transient changes to all its moment of 
inertia (MI) tensor components. This will directly affect its rotation and, hence, the pulse profiles. 
Considering that the measurements of pulse timings have reached extraordinary precision, even minute changes 
in the star's moment of inertia may be observable, providing a sensitive probe for QCD phases in the core of 
these dense objects. By analogy, asteroid-neutron star mergers should also cause a pulse profile modulation, 
offering a falsifiable consequence of the collision model for various transient phenomena. 

The primary purpose of this work is to investigate 
such a correlation by determining the effects of asteroid-neutron star collisions on irregularity in the pulsar's 
pulse profile, and help constrain the collision model of GRBs and FRBs.
The paper is organized in the following manner. 
Section \ref{section:sec2} briefly reviews the asteroid-neutron star collision model proposed by 
\citet{Colg81}, highlighting only those features that are relevant to the present work. In 
section \ref{section:sec3}, we study the impact of a generic perturbation in the MI tensor components 
on pulsar dynamics. Section \ref{section:sec4} is devoted to the computation and diagonalization of 
the perturbed MI tensor, from which the principal axes of the perturbed pulsar are determined. In section 
\ref{section:sec5}, we describe the numerical procedure used to compute the pulse-profile modulation, 
and present the results. Finally, section \ref{section:sec6} summarises our conclusions and discusses 
the implications of our findings.
\section{Asteroid-Neutron Star Collision and the Fate of the Asteroid}
\label{section:sec2}
A direct impact of an asteroid (or, could be a comet) of mass $\sim 10^{18}$ g on a neutron star could 
be a possible cause for the unique March 5, 1979 gamma burst as suggested by \citet{Colg81}. The authors 
considered the following sequence of events of the asteroid before and after the collision. 
As the body enters a strong gravitational field, the rapidly increasing tidal force elongates 
the body radially and flattens it between magnetic longitudes. Initially, the internal strength of 
the material resists radial compression, and the distortion proceeds approximately at constant volume. 
At sufficiently small radii, the tidal stress exceeds the cohesive strength of the solid, leading to 
fragmentation. For simplicity, the disruption is treated as a single breakup event, after which the 
subsequent dynamics of the infalling material are followed.

The fragmented material then encounters the intense dipolar magnetic field of the neutron star. Fast compression 
makes the fragments highly conducting, effectively as diamagnetic objects. The magnetic field guides infall 
by channelling material along field lines, and the fragments are focused onto a limited region of the stellar surface.
Impact with the neutron star's surface releases kinetic energy, driving shock heating and the rapid expansion of matter. 
This interaction gives rise to a transient, high-energy emission identified with the observed gamma-ray burst. The 
explosion at the impact site ejects material along connected magnetic flux tubes, while a fraction of the matter remains 
gravitationally bound. Following the initial burst, fallback and redistribution of the accreted material occur
in the conjugate magnetic field points.

The deposition of debris in the magnetic conjugate points  
are key elements of the \citet{Colg81} model, which serves as the basis for our study. 
As shown in Fig. \ref{fig:slab}, the fallback material is deposited onto the outer crust of the neutron star in the form of two 
nearly rectangular slabs ($A_1$ and $A_2$), each of length $l \simeq 2.5\ \mathrm{km}$, depth
$d \simeq 20\ \mathrm{m}$, and width $w \simeq 100\ \mathrm{m}$, with a mass density of order $10^6\ \mathrm{g~cm^{-3}}$. 
The slabs $A_1$ and $A_2$ correspond to magnetic conjugate points located at the same magnetic longitude and 
at polar angles $\delta$ and $(\pi-\delta)$, respectively, measured with respect to the magnetic axis.
We will calculate the effect of this deposition on pulsar's MI tensor in section \ref{section:sec4}.
\section{Effects of Moment of Inertia Perturbations on Pulsar Rotation}
\label{section:sec3}
Before the asteroid-neutron star collision, the (unperturbed) pulsar is taken as oblate 
spheroid, rotating about the symmetric $z_\Omega$-axis with angular frequency $\omega$ 
and angular momentum $L$ pointing along $\mathbf{\hat{z}_\Omega}$ (Fig. \ref{fig:slab}). 
The diagonal components of the MI tensor with respect to the principal axes ($x_\Omega, y_\Omega, z_\Omega$) 
are denoted by $I_{ij}^0$ ($i, j = 1,2,3$), conveniently written as $I_{11}^0 \equiv I_1^0$, 
$I_{22}^0 \equiv I_2^0$ with $I^0_1=I^0_2$ and $I_{33}^0 \equiv I_3^0 = I_0$ (with $I_0 > 
I^0_1, I^0_2$), and $I_{ij}^0 = 0$ for $i \ne j$. The star's oblateness $\eta = (I_0 - I_1^0)/I_0$ 
depends on the star's kinematics, such as mass, rigidity of the crust, and magnetic field, etc. 
The values of $\eta$ remain uncertain at present, whereas the theoretical simulations 
\citep{horowitz09, baiko18} constrain $\eta \sim 10^{-6}$, magnetar modeling allows as large as 
$\sim 10^{-4}$ \citep{maki14}. From an observational perspective, targeted searches for 
gravitational waves from isolated pulsars, assuming triaxiality, place limits of $\eta \lesssim 10^{-5}$ 
for the Crab and $\eta \lesssim 10^{-4}$ for Vela \citep{ligo20}. Some pulsars may even reach 
$\eta \sim 10^{-2}$-$10^{-3}$ as suggested by \citet{aasi14}. 

Following the merger, the asteroid material is deposited in the rigid crust of the pulsars, 
as described in the previous section. Eventually, all this material should get 
redistributed by equilibration processes. Our calculations apply to the period from the initial 
formation of these slabs from asteroid impact, until complete thermalisation and re-homogenisation 
occurs. We will assume that the asteroid material deposition process, in the form of 
two slabs, is instantaneous. This helps set the initial conditions for solving the Euler equations below.
The asymmetric deposition of the materials modifies all 
the moment of inertia tensor components $I_{ij}$ ($i, j = 1,2,3$). Thus, the perturbed MI tensor 
generates off-diagonal terms, inducing precession motion of the pulsar. The perturbed MI tensor 
components and the resulting new set of principal axes are determined numerically (see section
\ref{section:sec4}). Note that a pulsar may already have a wobbling motion. Our calculations 
then lead to additional wobbling. Here, for simplicity, we neglect any pre-existing wobbling by 
assuming a spheroid shape and spinning about its principal axis.
The rotational motion of the perturbed pulsar can be studied using Euler’s equations:
\begin{align}
I_1 \dot{\omega}_1 - (I_2 - I_3)\omega_2\omega_3 &= 0, \\
I_2 \dot{\omega}_2 - (I_3 - I_1)\omega_1\omega_3 &= 0, \\
I_3 \dot{\omega}_3 - (I_1 - I_2)\omega_1\omega_2 &= 0.
\end{align}

Here, $I_1, I_2, I_3$ are the principal moments of inertia and 
$\omega_1,\omega_2,\omega_3$ are the components of angular velocity 
about the principal axes.
Assuming the perturbed pulsar predominantly rotates about the $z_\Omega$-axis, i.e., 
$\omega_3 \approx \text{constant} = \omega$ and $\omega_1,\omega_2 \ll \omega_3$, 
the above set of coupled equations reduces to simple harmonic motion in $\omega_{1,2}$ 
(see \cite{Bagchi_2022}):
\begin{equation} \label{eq:Omega}
\ddot{\omega}_1 + \Omega^2 \omega_1 = 0, \qquad
\ddot{\omega}_2 + \Omega^2 \omega_2 = 0, 
\end{equation}
with the solutions, 
\begin{align}
\omega_1 (t) & =   \omega_1^0 \cos (\Omega~t) - \frac{\omega_2^0}{k}  
\sin (\Omega~t) \label{eq:omega10} \\
\omega_2 (t) & =  k \omega_1^0 \sin (\Omega~t) + \omega_2^0  
\cos (\Omega~t).\label{eq:omega20}
\end{align}
Where, $\Omega = \omega_3 \sqrt{\frac{(I_3 - I_1)(I_3 - I_2)}{I_1 I_2}}$ is the precession frequency, 
$k = \sqrt{\frac{I_1 (I_3 - I_1)}{I_2 (I_3 - I_2)}}$ and $\omega_1^0$, $\omega_2^0$ are two 
arbitrary constants to be determined using the conservation of angular momentum 
(see later in section \ref{section:sec4.2}).
\section{Perturbed MI Tensor and Principal Axes of the Pulsar}
\label{section:sec4}
\begin{figure}
\includegraphics[width=\columnwidth]{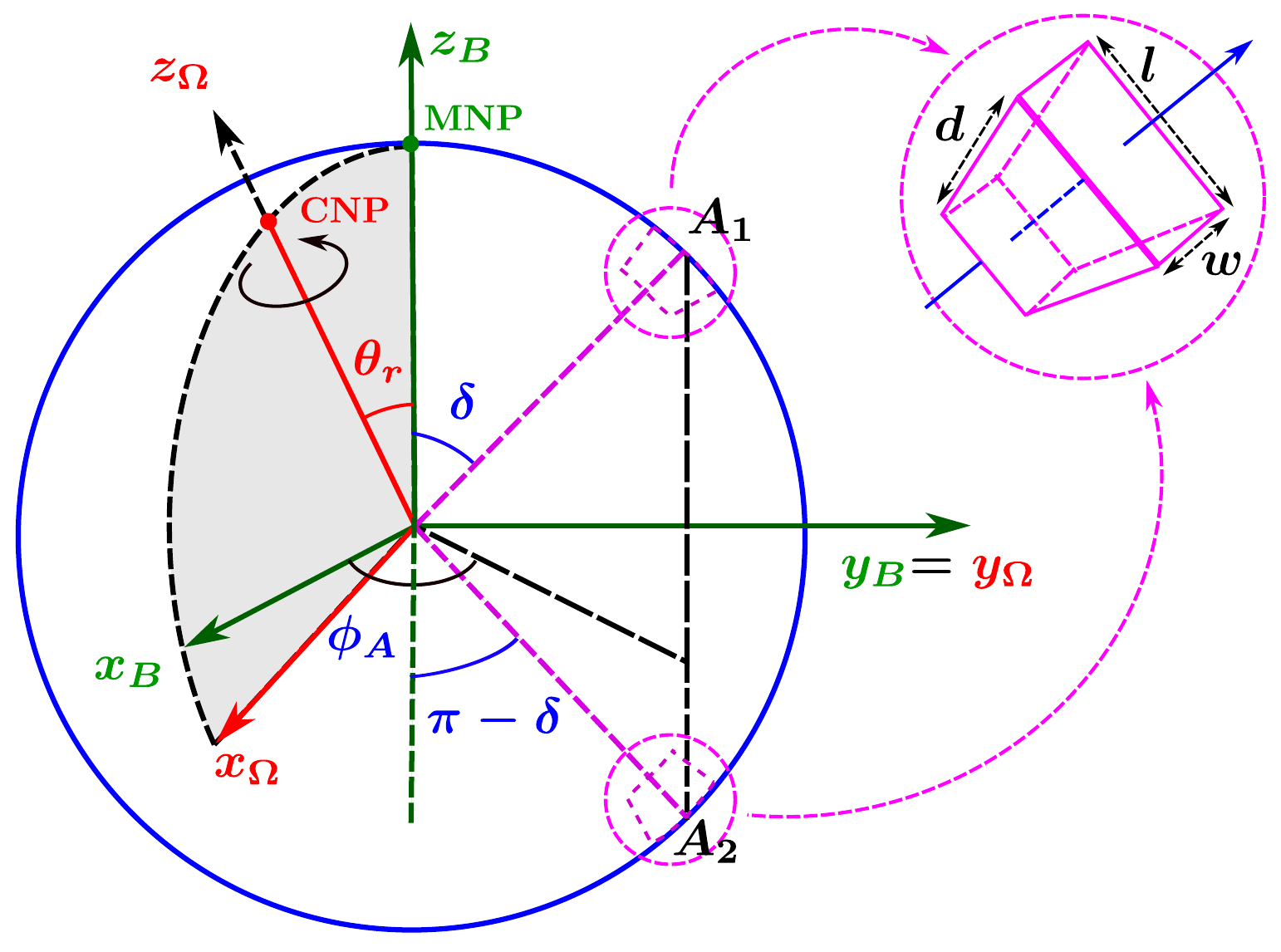}
\caption{ 
Before the asteroid-pulsar collision, the unperturbed pulsar rotates about
$\mathbf{\hat{z}_\Omega}$-axis with angular frequency $\omega$ and angular momentum 
$\vec {L} = L\mathbf{\hat{z}_\Omega}$. After collision, the asteroid material deposits in the form 
of two slabs, $A_1$ and $A_2$, each of dimensions $(d\times l\times w)$, located at polar angles $\delta$
and $(\pi-\delta)$ and azimuthal angle $\phi_A$, measured relative to the {\it magnetic} frame 
$S_B(x_B, y_B, z_B)$. $\theta_r$ is the angle between $\mathbf{\hat{z}_\Omega}$ and $\mathbf{\hat{z}_B}$.
MNP and CNP denote the Magnetic North Pole and the Celestial North Pole, respectively. }
\label{fig:slab}
\end{figure}
This section focuses on evaluating the components of the MI tensor of the pulsar resulting from 
the deposition of the asteroid materials. We will then numerically determine the 
resulting set of principal axes. For convenience, we shall refer to the frame $S_B(x_B, y_B, z_B)$ 
as the {\it magnetic} frame, and the principal axes $S_\Omega (x_\Omega, y_\Omega, z_\Omega)$ of  
the unperturbed pulsar as the {\it rotating} frame (see Fig. \ref{fig:slab}).
The directions $\mathbf{\hat{z}_B}$ 
(Magnetic North Pole) and $\mathbf{\hat{z}_\Omega}$ (Celestial North Pole) are uniquely specified. 
The other set of axes can be chosen at one's convenience, provided they remain a
right-handed rectangular system. Here, the $y_B$ axis is chosen to coincide with the $y_\Omega$ axis. 
The angle between the magnetic axis ($\mathbf{\hat{z}_B}$ ) 
and rotation axis ($\mathbf{\hat{z}_\Omega})$ is denoted by $\theta_r$. The slabs $A_1$ and $A_2$ 
are taken at the azimuthal angle $\phi_A$ with polar angles $\delta$ and $(\pi-\delta)$, respectively, 
relative to the magnetic frame. Since $S_B$ and $S_\Omega$ frames share a common $\mathbf{\hat{y}}$-axis 
(by our choice), their relative orientation is obtained through a rotation
$\mathbf{R_y}(\theta_r)$ about the $\mathbf{\hat{y}}$-axis by an angle of $\theta_r$:
\begin{equation} \label{eq:rzero}
    \mathbf{R_y}(\theta_r) = \begin{bmatrix}
        \cos \theta_r & 0 & -\sin \theta_r\\
        0 & 1 & 0\\
        \sin \theta_r & 0 & \cos \theta_r
    \end{bmatrix}
\end{equation}
For convenience, we first compute the MI tensor $I^B$ in the magnetic frame $S_B$, and 
subsequently transform it to the rotational frame $S_\Omega$ via,
    \begin{equation} \label{eq:miomega}
        I^\Omega = \mathbf{R_y}(\theta_r) I^B \mathbf{R_y}^{-1}(\theta_r).
    \end{equation}
\subsection{The Perturbed MI Components}
The MI tensor components in the magnetic frame of reference are given by 
(ignoring relativistic corrections),
    \begin{equation}
        I^B_{ab}= \int \rho(\vec r) (\delta_{ab}r^2 - r_a r_b)d^3r
    \end{equation}
As an illustration, the $I^B_{11}$ component for each slab, expressed in Cartesian coordinates, 
is given by   
\begin{equation}  \label{eq:micomp}
I^B_{11}= \iiint \rho(\vec r) ( y^2 + z^2) d^3 r.
\end{equation}
As mentioned earlier, the centers of mass of the slabs $A_1$ and $A_2$ are taken at azimuthal 
angle $\phi_A$ and polar angle $\delta$ and $(\pi-\delta)$ respectively relative to the magnetic
frame. When calculating the slab's contributions to the moments of inertia, we assume the star 
is spherical and neglect its intrinsic oblateness ($\eta$). This is justified since a slab of 
mass $\Delta m \simeq 10^{-15}\, M_\odot$, deposited at a polar angle $\theta$, contributes 
at the level $I^{\rm slab} \sim \Delta m R^2 \sin^2\theta$. The fractional contribution 
relative to the unperturbed MI is easily seen to be $I^{\rm slab}/I^0
\simeq (\Delta m/M_\odot) (1+2\eta) \simeq 10^{-15}(1+2\eta)$. 
Hence, the correction due to the stellar deformity can 
be safely ignored. Therefore, $I^B_{11}$ from Eq. (\ref{eq:micomp}) can be expressed in spherical 
coordinates as,
\begin{align}
I^B_{11}
&=
\iiint
\rho(r)\,
\bigl(
r^{2}\sin^{2}\theta\,\sin^{2}\phi
+
r^{2}\cos^{2}\theta
\bigr)~r^{2}\sin\theta
\, dr\, d\theta\, d\phi ,
\end{align}
where the limits of integration are $(R-d) \le r \le R$, 
$(\delta-\frac{l}{2R}) \le \theta \le (\delta+\frac{l}{2R})$, 
and $(\phi_A-\frac{w}{2R\sin\delta}) \le \phi \le (\phi_A+\frac{w}{2R\sin\delta})$.
Similarly, all other components are computed in the $S_B$ frame, and the resulting contributions 
from both slabs are added to the unperturbed MI tensor of the star. Finally, the MI tensor 
$I^\Omega$ is obtained using the transformation as given in Eq. (\ref{eq:miomega}).

\subsection{Determination of the Principal Axes and Their Relative Orientations}
\label{section:sec4.2}
\begin{figure}
\centering
\includegraphics[width=0.6\linewidth]{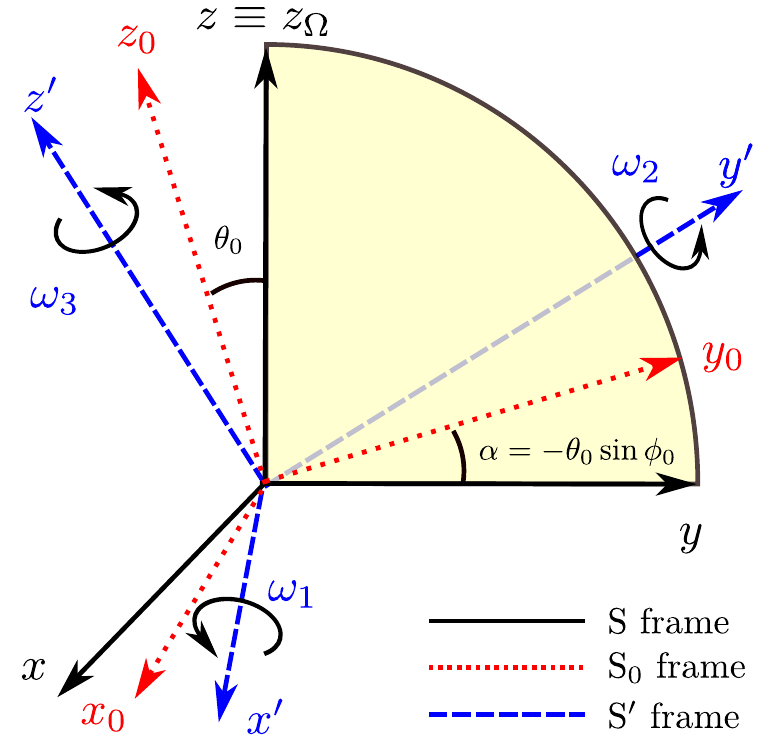}
\caption{$S(x, y, z)$ is the space-fixed frame (black solid lines) with respect to which the 
pulse profile is analyzed. The orientations 
of the principal axes $S_0(x_0, y_0, z_0)$ of the perturbed 
pulsar at $t=0$ relative to $S(x, y, z)$ are shown by red dotted lines. Blue dashed lines represent 
the body-fixed $S^\prime$ frame at an arbitrary time $t$ (see the text for details).}
\label{fig:axes}
\end{figure}
We follow the standard (diagonalization) prescription to determine the set of principal axes 
($x_0, y_0, z_0$), i.e., by finding the eigenvalues of the MI tensor 
$I^\Omega$ and the corresponding eigenvectors. For the analysis of pulse profile 
modulations, we chose a set of space-fixed frame $S(x,y,z)$ (Fig. \ref{fig:axes}) obtained from the frame 
$S_\Omega$ (Fig. \ref{fig:slab}) by a rotation about $z_\Omega$-axis such that $y_0$ lies in $y$-$z$ plane.
The reason for this choice will become clear as we proceed further. The 
orientations of the principal axes relative to the newly 
chosen $S(x,y,z)$ frame are related to the coordinate transformations parameterized by the angles 
$(\theta_0, \phi_0)$. Where $(\theta_0, \phi_0)$ is the set of polar and azimuthal angles of $z_0$, 
with respect to the $S$ frame. As argued in Ref. \citep{og22}, such 
transformations can be described by the rotation matrix,
\begin{equation}
R_0 = 
\begin{pmatrix} 
1 & 0 & -\theta_0 \cos\phi_0 \\
0 & 1 &  -\theta_0 \sin\phi_0 \\
\theta_0 \cos\phi_0 & \theta_0 \sin\phi_0 & 1 \label{eq:rot}
\end{pmatrix}
\end{equation}

With the above choice of $S$ frame, one can resolve the angular momentum 
$\vec L$ along the principal axes  
\begin{align}\label{eq:comp1}
L_{x_0} &= I_1 \omega_1^0 = - L \theta_0 \cos \phi_0 \\
L_{y_0} &= I_2 \omega_2^0 = - L \theta_0 \sin \phi_0 \\
L_{z_0} &= I_3 \omega_3^0 \simeq L.
\end{align}
Note that, in the collision model of \citet{Colg81}, the asteroid 
approaches the pulsar radially and therefore imparts no angular momentum. Accordingly, 
we assume the conservation of angular momentum in the collision.
Now, substituting $\omega_1^0$ and $\omega_2^0$ in Eqs. (\ref{eq:omega10}) and (\ref{eq:omega20}), 
we obtain
\begin{align}
\omega_1 (t)&=\dot\theta_1=-\omega \theta_0 [\cos \phi_0 \cos (\Omega t) 
- \frac{\sin\phi_0}{k} \sin(\Omega t)]\label{eq:soln1}\\
\omega_2 (t)&=\dot \theta_2 = -\omega \theta_0 [k \cos \phi_0 \sin (\Omega t) 
+ \sin\phi_0 \cos(\Omega t)]. \label{eq:soln2} \\
\omega_3 (t)&=\dot \theta_3 =\omega.\label{eq:soln3}
\end{align}
The above set of equations describes the rotational dynamics of the perturbed pulsar and 
allow us to determine their impact on the pulse profile, as discussed below. 
\section{Pulse-Profile Modulation: Numerical Method and Results} 
\label{section:sec5}
\begin{figure}
\centering
\includegraphics[width=0.6\linewidth]{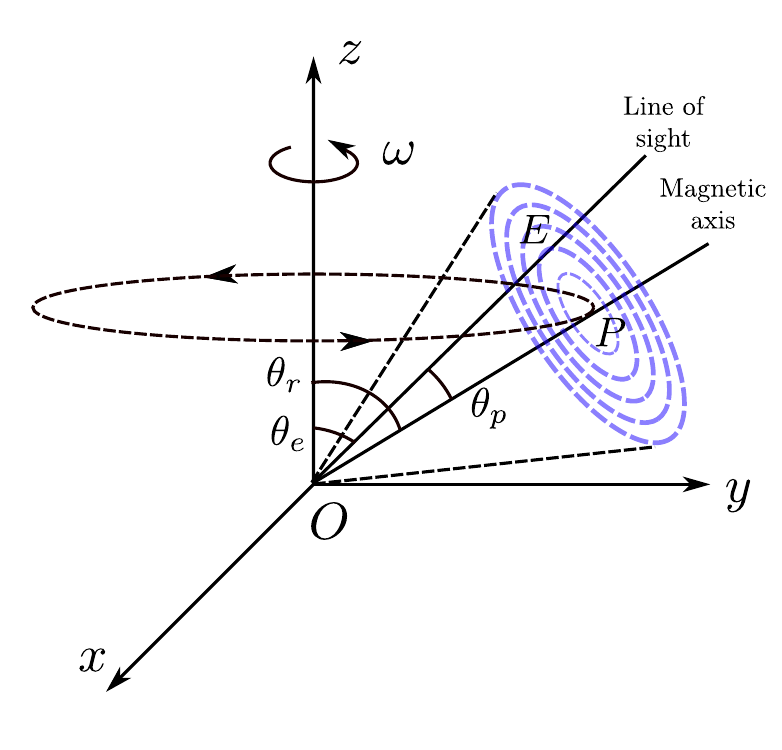}
\caption{Figure shows the radiation emission cone of a pulsar. The magnetic 
axis (OP) and the line of sight (OE) pointing towards Earth make an angle 
$\theta_r$ and $\theta_e$, respectively with the rotation axis. 
$\theta_p$ is the angle between OP and OE. [Taken from \citet{og22}.]}
\label{fig:conic}
\end{figure}

For the numerical analysis of the pulse profile, we follow the algorithm developed by some of 
us in Ref. \citep{og22} to study the effects of phase transitions on pulse-profile modulations. Here, 
we briefly summaries the salient features of the analysis. For the unperturbed pulse profile, we use the 
standard conical emission geometry \citep{gil81, gil84} with Gaussian pulse profile of width $\beta$:
\begin{equation} \label{eq:profile}
F(\theta_p) = F_0~e^{-(\theta_p^2/\beta^2)}.
\end{equation}
Here, $\theta_p(t)$ denotes the angle between $\vec{OP}$ and $\vec{OE}$, as shown in Fig.~\ref{fig:conic},
and $F_0$ is maximum value of the flux. 
The point $P$, representing the center of the pulse-emission region, and the point $E$, denote 
the intersection of the emission region with the line of sight (OE), both lie on the stellar surface. 
The magnetic axis $z_B$ (i.e., the axis of the conical emission) 
and the line of sight make angles $\theta_r$ and $\theta_e$, respectively, with the rotation axis 
$z_\Omega$. The angle $\theta_p(t)$ varies as the emission cone sweeps across the observer with the 
rotational frequency $\omega$. Pulsar's precession caused by the perturbation modifies the temporal 
evolution of $\theta_p(t)$ and hence, modulates the observed pulse profile. The evolution of $\theta_p(t)$ 
is obtained stepwise by integrating Eqs.~(\ref{eq:soln1})--(\ref{eq:soln3}). The initial (i.e., at $t=0$) 
orientations of $S_0$ relative to $S$ (Fig. \ref{fig:axes}) is known through $R_0$ (Eq. \ref{eq:rzero}). 
Thus, the point $P(\theta_r,\phi_r)$ in the body fixed frame at time $t = 0$ is obtained by $R_0 [(\theta_r,\phi_r)]$.  
As the perturbed pulsar evolves, the angular displacement $\theta_i (\Delta t)$, ($i = 1,2,3$) is computed 
by integrating Eq. (\ref{eq:soln1}) - Eq. (\ref{eq:soln3}) for a time step $\Delta t$.  The matrix 
$R_1$ describing the orientations of say, $S_1$-frame (the body-fixed frame at $\Delta t$) relative 
to the $S_0$-frame is then obtained through $R_1 = R_x (\theta_1)R_y (\theta_2)R_z (\theta_3)$. 
The coordinates of $P$ at $t = \Delta t$ as seen by $S$-frame is determined by the transformations, 
$[\theta (\Delta t), \phi (\Delta t)] = R_0^{-1} R_1^{-1} R_0 (\theta_r, \phi_r)$. We then calculate 
$\theta_p (\Delta t)$, and hence the flux of the pulse profile $F(\theta_p)$ using 
Eq. (\ref{eq:profile}). The above procedure is followed sequentially for the successive time-steps 
$2 \Delta t, 3 \Delta t, ...$, etc., for a sufficiently long duration to observe the imprint on
pulse profile. Stable pulse-profile generation is ensured by choosing a small time step 
$\Delta t = 10^{-8}$ s.

\begin{figure}
\centering
\includegraphics[width=1.0\linewidth]{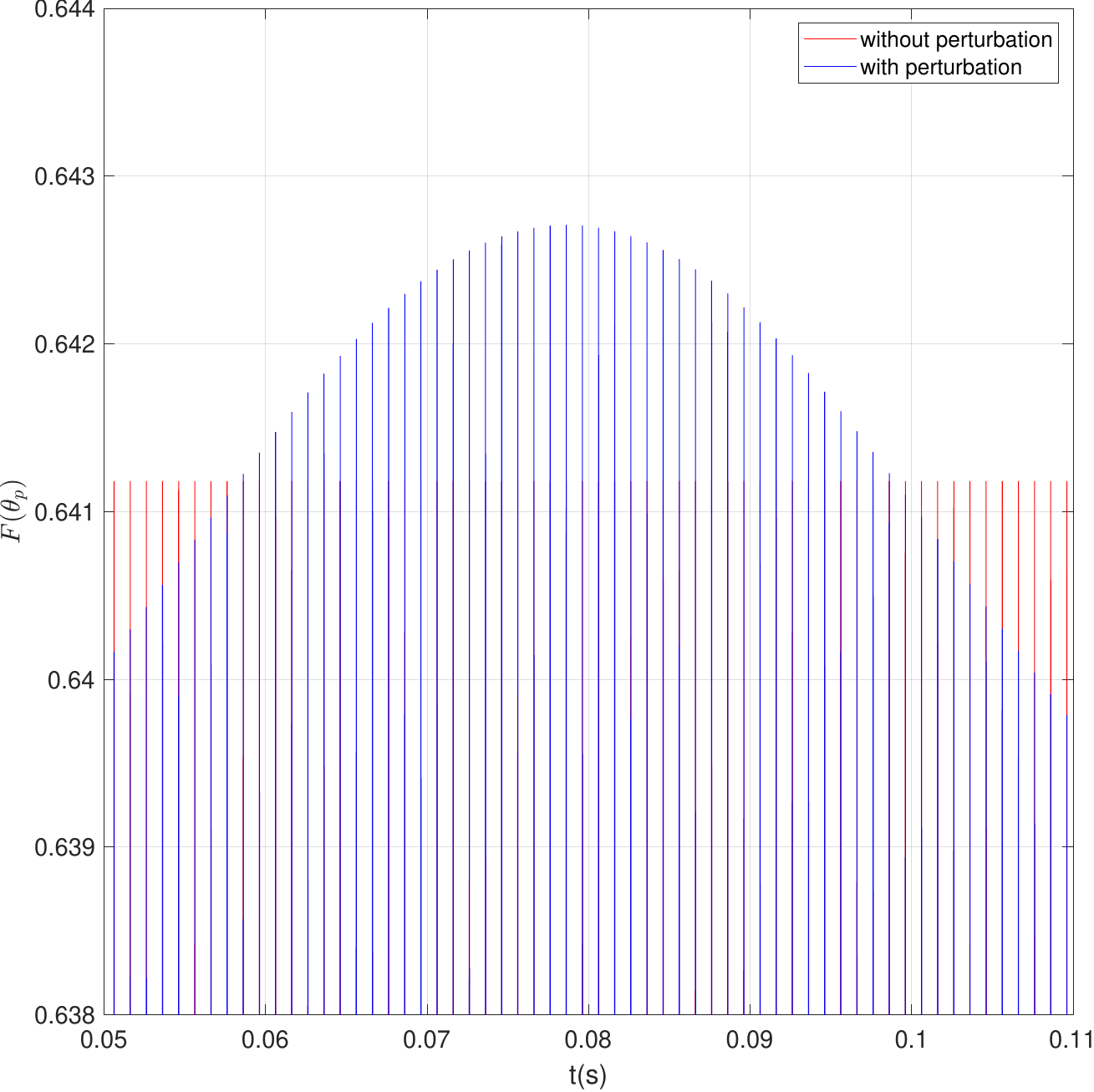}
\caption{Temporal evolution of the normalised ($F_0 = 1$) flux $F(\theta_p)$ for a millisecond 
pulsar without wobbling (red) and with wobbling (blue). We take $\delta = 60^\circ$, 
$\phi_A  = 45^\circ$ for the wobbling-induced motion and adopt comparatively large values 
of $\epsilon = 10^{-5}$ and $\eta = 10^{-2}$.}
\label{fig:fig4}
\end{figure}
\begin{figure}
\centering
\includegraphics[width=1.05\linewidth]{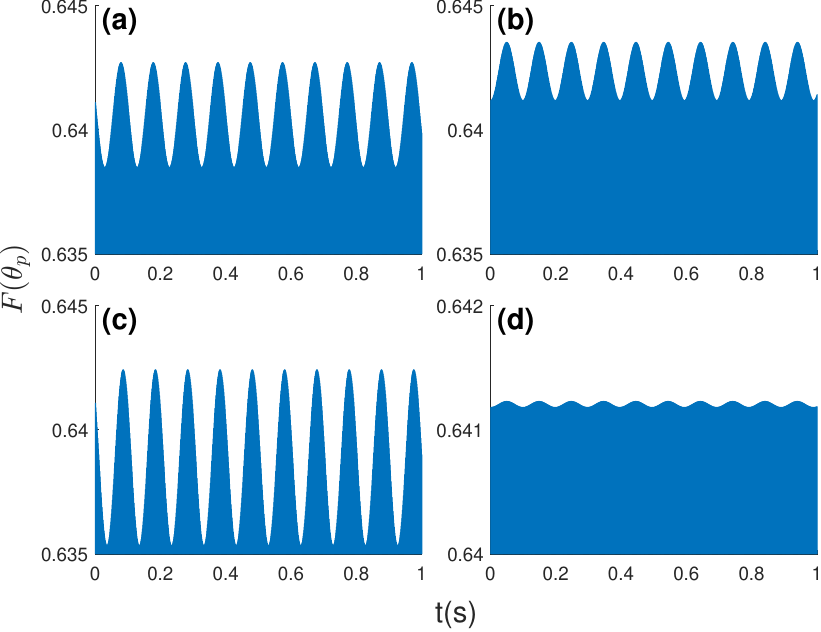}
\caption{Time evolution of the flux $F(\theta_p)$ (with $F_0 = 1$), exhibiting a characteristic 
modulation timescale $T_\Omega \simeq 0.1 \mathrm{s}$ arising from perturbations induced by a 
neutron star-asteroid collision. For computational limitation, we consider relatively large values 
of $\epsilon (= 10^{-5})$ and $\eta (= 10^{-2})$. 
Top panels : (a) $\delta = 60^\circ$, $\phi_A  = 45^\circ$; (b)  
$\delta = 60^\circ$, $\phi_A  = 90^\circ$. Bottom panels : 
(c) $\delta = 90^\circ$, $\phi_A = 45^\circ$; (d) $\delta = 90^\circ$, $\phi_A = 90^\circ$. 
Note that the pulse profile is nearly unaffected in panel (d).} 
\label{fig:fig5}
\end{figure}
\begin{table}
    \centering
\caption {The parameters used for simulations: 
$\delta$ denotes the polar angle of one slab with respect to the magnetic axis (see Fig.~\ref{fig:slab}), 
with the other located at $(\pi-\delta)$. Multiple azimuthal angles ($\phi_A$) are considered for each 
value of $\delta$ to observe the azimuthal dependence. The last column shows the approximate values of the 
angle $\theta_0$ (see Fig.~\ref{fig:axes}) 
through which the new principal axis $z_0$ (obtained via diagonalization) is tilted with respect to the $z$-axis 
due to an asteroid collision.}
    \label{tab:cases}
    \begin{tabular}{l|c|c}
     \hline 
          $\delta$ & $\phi_A$ & $\theta_0$ (in radian)\\
    \hline
        $60\degree$ & $45\degree$  & $6.5 \times 10^{-4}$ \\
        $60\degree$ & $90\degree$ &  $3.6 \times 10^{-4}$ \\
        $90\degree$ & $45\degree$ & $ 10^{-3}$ \\
        $90\degree$ & $90\degree$ & $7.4 \times 10^{-6}$     \\   
    \hline
    \end{tabular}
\end{table}
Before presenting the results of the numerical simulations, we briefly review, following the arguments 
of Ref. \citep{og22}, the analytical estimates of the quantities relevant to flux modulation induced 
by wobbling. 
We parametrize the perturbation in the moment of inertia tensor as $(\Delta I / I) \simeq \epsilon$. 
For an asteroid of mass $m$, the perturbation may be expressed as $\epsilon \simeq (m / M_\odot) \simeq 
10^{-15}$. 
For $\epsilon$ small compared to the pulsar's oblateness $\eta$, i.e.  
$\epsilon \ll \eta$, the precession frequency $\Omega = [(I_3 - I_1)(I_3 - I_2)/(I_1 I_2)]^{1/2} \omega$ 
(see Section \ref{section:sec2}), reduces to $\Omega \simeq \eta \omega$, corresponding to a
flux modulation timescale $T_\Omega \simeq T_\omega / \eta$. Where, $T_\omega$ is the pulsar spin period. 
Thus, for an asteroid colliding with a spheroidal pulsar of oblateness $\eta \simeq 10^{-6}$, 
the resulting flux modulation becomes appreciable only after of order $10^{6}$ rotations. In contrast to the analytical
estimate of precession frequency, quantifying the magnitude of the 
modulated flux relative to the unperturbed case is not straightforward due to the complexity involved in 
the wobbling motion. Nevertheless, an order of magnitude estimate of the fractional flux variation may be 
obtained from Eq. (\ref{eq:profile}) as $(\delta F/F) \simeq (2\theta_p/\beta^2) \delta \theta$. 
Substituting $\theta_p \simeq 10^\circ$, $\beta = 15^\circ$, and taking $\delta \theta \simeq \theta_0$ 
to account for the maximum effect, we obtain $(\delta F/F) \simeq  5 \theta_0$.

The parameter values adopted in our simulations are summarized in Table \ref{tab:cases}.
The angle between the magnetic and rotation axes (see Fig.~\ref{fig:conic}) 
is taken at $\theta_r = 30^\circ$ (observational estimates for a large sample of pulsars show
a range around $30^\circ$ \citep{Malo90}), while the line of sight is assumed to be inclined
at an angle $\theta_e = 40^\circ$ with respect to the rotation axis. 
Due to the computational constraints, we employ relatively large values of $\eta = 10^{-2}$ 
and $\epsilon = 10^{-5}$. These values yield a flux modulation timescale of order 
$\sim 0.1$ s for a millisecond pulsar. A magnified view of the time evolution (choosing a shorter time 
interval) of the flux $F(\theta_p)$ is shown in Fig.~\ref{fig:fig4}. Each vertical line represents an 
individual pulse of a millisecond pulsar.The effect of wobbling induced by an asteroid collision manifests 
as variations in the flux amplitude relative to that of the unperturbed pulsar.
The maximum fractional change $(\delta F/F)$, of order 
$\simeq 10^{-3}$, is also found to scale approximately with $\theta_0$ (see Table \ref{tab:cases} for the value of
$\theta_0$). The pulse profiles on relatively longer timescale across all panels in Fig.~\ref{fig:fig5} exhibits the 
flux modulations with a characteristic timescale $T_\Omega \simeq T_\omega/\eta \simeq 0.1 \mathrm{s}$, 
consistent with the analytical estimate. For $\delta = \phi_A = 90^\circ$, the pulse profile remains nearly 
unchanged, as the off-diagonal components of the moment of inertia tensor are found to be negligible, 
and hence do not induce significant wobbling. Note that, barring the dependence on the magnitude of 
the perturbation ($\epsilon$), the relative flux modulation also depends on distributions of the 
asteroid materials. Thus, the temporal variation of the flux provides complete information 
about the characteristics of the perturbations and the pulsar's oblateness.

Before concluding this section, we make a brief comment on our earlier prediction in Ref.  
\citep{og22} regarding the possibility of an additional characteristic (\textit{second}) 
timescale associated with the flux modulation. A simplified 
analytical estimate suggested that such a timescale $T_m \simeq (\epsilon/\eta) T_\omega$, depends 
on both $\eta$ and $\epsilon$, and thus encodes information about the perturbation $\epsilon$ 
(in contrast to $T_\Omega$, which depends solely on the pre-existing oblateness $\eta$).
We are not reporting any results here for a second modulation due to numerical 
uncertainties. Note that, as per the above estimate, the second modulation should be on a much 
larger time scale. Various numerical errors can accumulate to become significant when integrating 
over very large time scales, e.g., the one arising from applying various rotation transformations 
with discrete time steps (even though we use very small time steps) and the resulting noncommutativity. 
We hope to investigate this intriguing possibility of any other modulation time scale by a detailed, 
careful analysis in future.
\section{Conclusions and Discussion}
\label{section:sec6}
We adopt the pulsar-asteroid collision model of \citet{Colg81} and investigate the consequences of 
such impacts, focusing on potential imprints in observed pulse profiles. We find that the accretion 
of asteroid material onto the pulsar crust perturbs the moment-of-inertia tensor, thereby influencing 
its rotational dynamics. Changes to the diagonal components of the MI tensor can modify the pulsar's 
spin frequency at order $\epsilon$. However, such variations are unlikely to be directly detectable 
with current pulsar-timing precision. In contrast, the generation of off-diagonal components can 
induce precessional motion, leading to characteristic modulations in the pulse profile on a 
timescale enhanced by a factor of $1/\eta$ compared to the pulse timing.
Importantly, even small changes in the MI components, of order $\epsilon$, can produce large pulse 
profile modulations of order $\epsilon/\eta$ (depending on the relative location of asteroid material deposition).
Thus, if pulsar-asteroid collisions are responsible for phenomena such as GRBs and FRBs, the resulting 
correlated features in pulse profiles could provide a falsifiable test of these collision models.

We mention that the present analysis uses the algorithm developed by \citet{og22}, which is valid for 
small perturbations and is otherwise generic. This opens the possibility of exploring additional perturbations
that modify the pulsar's MI tensor, especially those that produce off-diagonal components, impacting the 
pulse profile. Such investigations are especially important in light of multiple observational indications 
of pulsar precession reported in the literature.

\section*{Acknowledgments}
AMS acknowledges support from the Raja Ramanna Chair position (DAE, Government of India).
The computations were performed on the {\it Chandra} server at BITS - Pilani (Pilani Campus).

\section*{Data Availability}
The data are not publicly available. The data are available from the authors upon 
reasonable request.
\bibliographystyle{mnras}
\bibliography{asteroid2026} 

\bsp	
\label{lastpage}
\end{document}